\documentclass[doublecol]{epl2}  
\usepackage[english]{babel}
\usepackage[latin1]{inputenc}
\usepackage{microtype}
\usepackage{palatino}
\usepackage{amssymb,amsthm,mathrsfs,amscd}
\usepackage[nointlimits]{amsmath}
\usepackage{cite}

\usepackage{url}
\usepackage{graphicx,color}

\newcommand{\ham}{\hat{\mathcal{H}}} 
\newcommand{\diff}{\mathrm{d}} 
\newcommand{\ii}{\mathrm{i}} 
\newcommand{\abs}[1]{\lvert#1\rvert} 
\newcommand{\ang}[1]{\langle#1\rangle} 
\DeclareMathOperator{\e}{e} 

\title{Dynamical decoupling and dynamical isolation in temporally modulated coupled pendulums}
\shorttitle{Dynamical decoupling and isolation in modulated coupled pendulums} 

\author{Grazia Salerno \and Iacopo Carusotto}
\shortauthor{G. Salerno {\em et al.}}
\institute{ INO-CNR BEC Center and Dipartimento di Fisica, Universit\`a di Trento, via Sommarive 14, 38123 Povo, Italy}

\abstract{We theoretically study the dynamics of a pair of coupled pendulums subject to a periodic temporal modulation of their oscillation frequency. Inspired from analogous developments in quantum mechanics, we anticipate dynamical localization and dynamical isolation effects, as well as the occurrence of non-trivial coupling phases. Perspectives in the direction of studying synthetic gauge fields in a classical mechanics context are outlined.}

\begin{document}
\maketitle

\section{Introduction}

Dynamical localization is a surprising consequence of quantum mechanics applied to particles subject to a strong time-dependent external force. This effect was first observed as renormalization of the magnetic response of an atom illuminated by a strong rf field~\cite{CTH}. In a solid state context, dynamical localization was proposed in~\cite{Dunlap,Ignatov,Zhao,Holthaus,Holthaus2} as a dramatic suppression of the d.c. conductivity of a metal in a strong a.c. field. Another closely related effect is the coherent destruction of tunneling in a double-well geometry, first predicted in~\cite{Grossmann, Grifoni} and extensively compared to dynamical localization in~\cite{DLvsCDT}.

While the experimental study of these effects in solids is made difficult by the unavoidable material imperfections and electronic decoherence, the robust coherence and the clean periodic potential experienced by atomic matter waves in temporally modulated optical lattices has allowed for a clear  observation of Bloch band suppression in a new atomic physics context~\cite{Madison}. Further studies of dynamical matter wave localization effects were reported in~\cite{Lignier,Arimondo09} using Bose-condensed atomic samples. Following the proposal~\cite{EckardtHolthaus, EckardtHolthausEPL}, this research line culminated in the observation of a dynamically-induced superfluid to Mott-insulator transition~\cite{Zenesini09}. 

Very exciting further developments of these ideas aim at using more complex modulation schemes to generate non-trivial hopping phases between the lattice sites~\cite{Sols08,Struck2012} and then synthetic gauge fields for neutral atoms~\cite{Kolovsky, CommentKolovsky, Hauke, Struck2013, Aidelsburger, Miyake}. Correspondingly to these advances in atomic physics, the same ideas are being explored in photonics to observe dynamical localization of light in coupled optical waveguides~\cite{Longhi} and, very recently, to generate synthetic gauge field for photons~\cite{Rechtsman, Hafezi}.

In this Letter, we report a theoretical study of dynamical localization phenomena in a classical mechanics context. The dynamic stabilization of the inverted pendulum when its pivot point is made to oscillate in space is a well celebrated example of non-trivial mechanical effect stemming from a temporal modulation of the system parameters~\cite{Butikov}. Here we consider a system of two coupled pendulums, whose oscillation frequencies are independently and periodically varied in time. In analogy to the coherent destruction of tunneling of a quantum particle in a double-well potential, we predict a {\em dynamical decoupling} effect, where exchange of energy between the pendulums is suppressed. When the pendulums are driven by an external force, we anticipate a novel {\em dynamic isolation} effect, where the temporal modulation effectively decouples the system from the external force. 

\section{The system and the theoretical model}

\begin{figure}[t]
	\centering
	\includegraphics[scale=0.35]{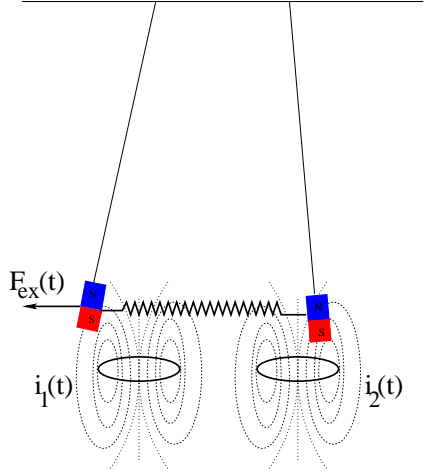}
	\caption{Sketch of the physical system under consideration. Each pendulum contains a magnet and the modulation of its natural oscillation frequency is controlled by the time-dependent current $i_{1,2}(t)$ flowing in the corresponding coil. The first pendulum may be externally driven by a time-dependent force $F_{\rm ex}(t)$.
	}
	\label{fig:sketch}
\end{figure}
The system of two identical coupled pendulums is modelled as a pair of coupled harmonic oscillators of equal masses $m$ following the motion equations:
\begin{align}
&m\, \dot{x}_1= p_1\label{eq1}\\
&\dot{p}_1=-m\omega_0^2\left[1+\nu_1(t)\right]\, x_1 +k (x_2-x_1)-\xi_1 \dot{x}_1\label{eq2}\\
&m\, \dot{x}_2= p_2\label{eq3}\\
&\dot{p}_2=-m\omega_0^2\left[1+\nu_2(t)\right]\, x_2 +k (x_1-x_2)-\xi_2 \dot{x}_2~. \label{eq4}
\end{align}
The $x_{1,2}$ variables indicate the spatial displacement of the pendulums from the equilibrium position. The linearised form of the motion equations is legitimate in the small oscillation regime where the displacements are much smaller than the length $L$ of the pendulums: in this regime, the natural oscillation frequency of each isolated pendulum is the usual $\omega_0=\sqrt{g/L}$. The friction constants of the two pendulums have the same value $\xi_1=\xi_2=\xi$. The coupling between the pendulums occurs via a spring of constant $k$.
The key element to achieve the dynamical decoupling and isolation effects is the temporal modulation of the system, which is included in Eq.~\eqref{eq2} and Eq.~\eqref{eq4} as a  temporal modulation of the restoring force strengths of relative amplitude $\nu_{1,2}(t)$.
One of the possible concrete realizations of this model is sketched in Fig.~\ref{fig:sketch}: each pendulum contains a magnet which feels the magnetic field generated by a coil located below its axis. In this way, the gravitational restoring force felt by each pendulum is supplemented by a contribution of magnetic origin, which can be controlled via the (time-dependent) current $i_{1,2}(t)$ flowing in the corresponding coil. The effective modulation of the natural oscillator frequencies then has the form:
\begin{equation}
 \omega^2_{1,2}(t)=\omega^2_0\,\left[1+\nu_{1,2}(t)\right],
 \label{omega}
\end{equation}
where the $\nu_{1,2}(t)$ are proportional to the currents $i_{1,2}(t)$. Introducing, for notational simplicity, the rescaled quantities:
\begin{equation*}
v_{1,2}(t)=\frac{\nu_{1,2}(t)\, \omega_0}{2}, \quad \Omega=\frac{k}{m\omega_0}, \quad \gamma=\frac{\xi}{2m},
\end{equation*}
the four equations of motion Eqs.~\eqref{eq1}-\eqref{eq4} can be summarized in a pair of complex equations:
\begin{equation}
\begin{split}
\dot{\alpha}_i&=-\ii\, \omega_0 \,\alpha_i - \ii \,v_i(t) (\alpha_i+\alpha_i^*)-\gamma\, (\alpha_i-\alpha_i^*)\\ &\quad\,+\ii\, \frac{\Omega}{2}\, (\alpha_{3-i}+\alpha_{3-i}^*-\alpha_i-\alpha_i^*). \label{alpha1}
\end{split}
\end{equation}
for the $\alpha_{i=1,2}$ complex variables defined as:
\begin{equation}
\alpha_i=\sqrt{\frac{ m \omega_0}{2}}x_i+ \ii \,\sqrt{\frac{1}{2 m \omega_0}} p_i.
\label{eq5}
\end{equation}
Complex conjugate equations hold for the $\alpha^*_{1,2}$. Physically the square modulus $|\alpha_i|^2$ is the instantaneous energy of the $i$-th pendulum and the argument of $\alpha_i$ is the oscillation phase. In experiments, both can be extracted by measuring the instantaneous position and velocity of the pendulums.

In order for the coupling between the oscillators to be effective, the strong coupling condition, $\Omega\gg \gamma$, will be assumed. Among the many forms of the modulation considered in the literature \cite{Sols08, Struck2012}, in the following we shall concentrate our attention on sinusoidal ones:
\begin{equation}
v_i(t)=(-1)^{i} I_0 \sin(w t).
\label{modulaz}
\end{equation}
As we shall see better in the following of this Letter, the most convenient regime where to obtain and observe the dynamical localization physics is characterized by the inequality chain:
\begin{equation}
\omega_0\gg w\gg\Omega.
\label{ineq}
\end{equation}
The latter inequality is essential to describe the system in terms of an effective, time-averaged coupling. The former helps to reduce those parametric instabilities that might otherwise occur for large values of the modulation amplitude $I_0$~\cite{Arnold, Berthet}. 

\section{Dynamical decoupling of the oscillators}

\begin{figure}[t]
	\centering
	\includegraphics[scale=0.67]{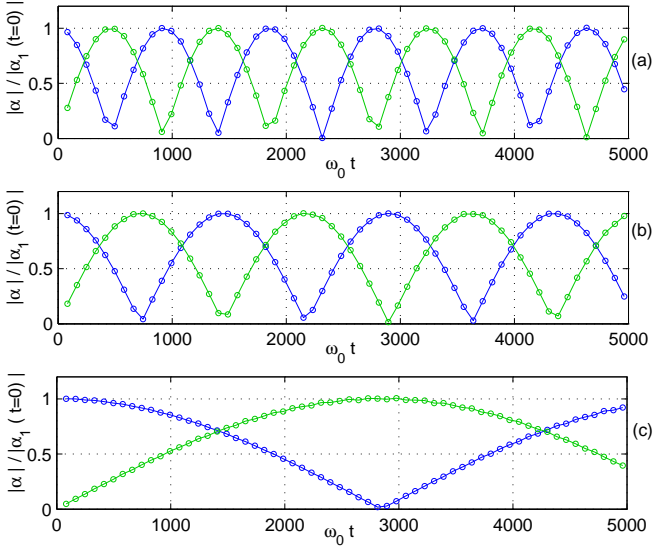}
	\caption{Numerical integration of the equations of motion Eq.~\eqref{alpha1}. The blue (green) dots indicate the modulus $\abs{\alpha_{1}}$ ($\abs{\alpha_{2}}$) of the oscillation amplitude of the first (second) pendulum, normalized to the initial amplitude, while the lines are guides to the eyes.  At the initial time $t=0$, only the first oscillator is excited, while $\alpha_2(t=0)=0$. The evolution of the two pendulums is stroboscopically followed at the modulation frequency $w$. The three panels are for different values of the modulation amplitude, $I_0/w=0$ (a), $I_0/w=0.632$ (b), and $I_0/w=1.053$ (c). System parameters are $w/\omega_0=7.6\times10^{-2}$, $\,\Omega/\omega_0=0.68\times 10^{-2}$, $\gamma/\omega_0=\,f_\text{ex}/\omega_0=0$. 
	}
	\label{fig:confrontoomegaeffnorwa}
\end{figure}

As a preliminary step, we have numerically integrated Eq.~\eqref{alpha1} in the vanishing friction case, $\gamma=0$. The integration method is a standard fourth order Runge-Kutta. In Fig.~\ref{fig:confrontoomegaeffnorwa} we have plotted the evolution of the $\abs{\alpha_{1,2}}$ sampled at the frequency $w$ of the modulation. At these stroboscopic times $t_j=2\pi j /w$, the system is described by a time-independent discrete evolution law. The different panels of Fig.~\ref{fig:confrontoomegaeffnorwa} correspond to the same modulation frequency $w$ but different values of the amplitude $I_0$.

In panel~(a), there is no modulation, $I_0=0$: the amplitudes of the two oscillators exhibit the usual beating effect, \textit{i.e.} a periodic exchange of energy between the oscillators at the coupling frequency $\Omega$. As a result of the modulation $v_i(t)$, for increasing values of its amplitude $I_0$, the frequency of the beat is modified: in particular, for the parameters of panels (b) and (c), the beat frequency is more and more reduced. For even larger amplitudes $I_0$, a non-monotonic behaviour of the effective beat frequency is observed (not shown). 

\section{Experimental proposal and dynamical isolation}
In order to perform a quantitative study of the effect of the modulation and, at the same time, to propose a viable procedure to experimentally observe these phenomena, it is useful to consider the realistic case of dissipative pendulums driven by an external force, which is assumed to be monochromatic, $F_\text{ex}(t)=2 f_\text{ex} \cos{\omega_\text{ex} t}$ and to act on the first pendulum only. Energy transfer to the second pendulum is made possible by the spring that couples the two pendulums.
The driven-dissipative motion equations for $\alpha_{i=1,2}$ then take the form:
\begin{equation}
\begin{split}
\dot{\alpha}_i&=-\ii\, \left(\bar{\omega}_0-\ii\gamma \right)\alpha_i - \ii \,v_i(t) (\alpha_i+\alpha_i^*)+\ii\, \delta_{i,1}F_\text{ex}(t)\\ &\quad-\ii\left(\frac{\Omega}{2}+\ii \gamma\right)\,\alpha_i^* +\ii\, \frac{\Omega}{2}\, (\alpha_{3-i}+\alpha_{3-i}^*)\label{nascostaalpha1noRWA}
\end{split}
\end{equation}
where we have defined the short-hand $\bar{\omega}_0=\omega_0 +{\Omega}/2$ and $\delta_{i,j}$ is the usual Kronecker delta.

\begin{figure}[t]
	\centering
	\includegraphics[scale=0.67]{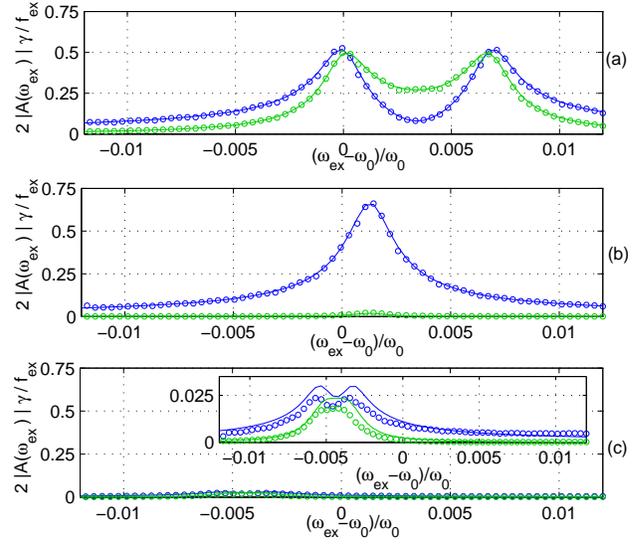}
	\caption{Response spectra $\abs{A_i(\omega_\text{ex})}$ as a function of the frequency $\omega_{\rm ex}$ of the external force. The stroboscopically sampled results are shown for the first (second) pendulum as blue (green) dots, normalized to the peak amplitude $f_{\rm ex}/(2\gamma)$ of a single isolated pendulum in the small $\gamma$ limit. The solid lines show the result of an analytical calculation based on the rotating wave approximation (see text). The different panels corresponds to different values of  the modulation amplitude, $I_0/w=\,0$ (a), $I_0/w=1.21$ (b), $I_0/w=2.31$ (c). The inset in (c) shows an enlargement of the main plot. System parameters: $w/\omega_0=7.6\times10^{-2}$, $\,\Omega/\omega_0=0.68\times 10^{-2}$, $\,\gamma/\omega_0=0.1\times10^{-2}$, $\,f_\text{ex}/\omega_0=4\times10^{-2}$. 
	}
	\label{fig:resonanceNORWA}
\end{figure}

For each $\omega_{\rm ex}$, the equations of motion Eq.~\eqref{nascostaalpha1noRWA} have been integrated until a steady-state regime, showing regular periodic oscillations, is achieved at long times $t\gg 1/\gamma$. For the stroboscopic sampling at $t_j=2 \pi j / w$, the steady oscillations have the form: $\alpha_i(t)\approx A_i(\omega_\text{ex}) \e^{-\ii\omega_\text{ex} t}$.

The complex amplitudes $A_i$ are obtained via a Fourier transformation of the stroboscopically sampled numerical solutions. The moduli $|A_i|$, as a function of the frequency of the external force, give the response spectra shown in Fig.~\ref{fig:resonanceNORWA}: each panel  corresponds to different value of the modulation amplitude $I_0$ and illustrates a different regime.

The case of no modulation is shown in panel (a): the spectra are characterized by a pair of peaks, split by $\Omega$ and of equal width $\gamma$. The lower (upper) frequency peak corresponds to the eigenmode where the two pendulums oscillate with the same (opposite) phase. At all frequencies, the oscillation amplitudes of the two pendulums remain comparable.

Panel (b) shows a case where the coupling of the two pendulums is dramatically suppressed: this effect is apparent in the figure as the two peaks merge into a single peak and no significant excitation is transferred to the second pendulum, which remains basically at rest with a negligible oscillation amplitude. This behaviour is the driven-dissipative manifestation of the {\em dynamical decoupling} effect, already seen in the lowest panel of Fig.~\ref{fig:confrontoomegaeffnorwa} for vanishing friction.

Panel (c) shows a novel regime: while some effective coupling of the two pendulums is still present, their global excitation by the external force is suppressed. The suppressed excitation is visible as a very small oscillation amplitude of both pendulums. The presence of a significant coupling is apparent in the inset where the response of the first pendulum is still showing a doublet.

\begin{figure}[t]
	\centering
	\includegraphics[scale=0.69]{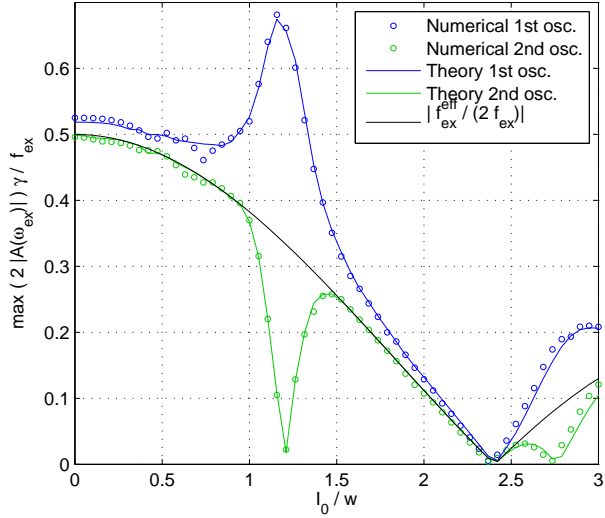}
	\caption{Blue (green) dots show the normalized maximum of the oscillation amplitude for the first (second) oscillator as a function of the modulation amplitude $I_0/w$. The solid lines show the result of an analytical calculation based on the rotating wave approximation (see text). System parameters as in Fig.~\ref{fig:resonanceNORWA}.}
	\label{fig:Feffettiva}
\end{figure}

These different regimes are illustrated in more detail in Fig.~\ref{fig:Feffettiva}. Response spectra have been numerically calculated for a number of different values of the modulation amplitude $I_0$. For each of these values, we plot the maximum of the amplitude $\abs{A_i(\omega_\text{ex})}$ over the external drive frequency $\omega_\text{ex}$, that is the resonant response at the peaks. The dynamical decoupling seen in Fig.~\ref{fig:resonanceNORWA}(b) corresponds here to the minimum of $\max(\abs{A_2})$ that is visible around $I_0/w\simeq 1.2$. The dynamical isolation seen in Fig.~\ref{fig:resonanceNORWA}(c) lies in the vicinity of the simultaneous minima of both $\max(\abs{A_{1,2}})$ that are visible around $I_0/w\simeq 2.4$. In the figure, note the further dynamical decoupling point around $I_0/w\simeq 2.7$. The solid lines are the analytical predictions of the rotating wave approximation that will be presented afterwards.

\begin{figure}[t]
	\centering
	\includegraphics[scale=0.77]{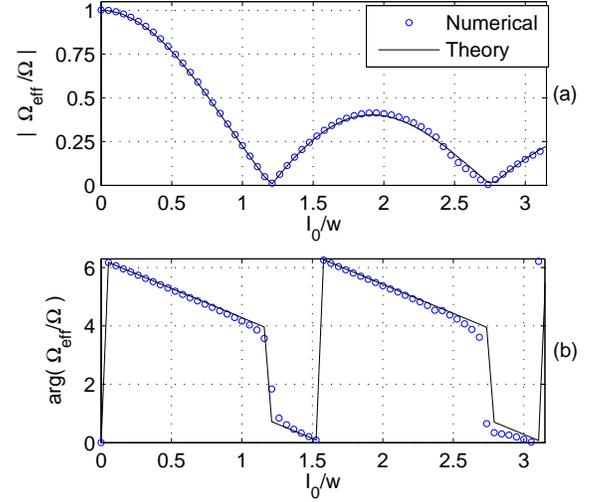}
	\caption{Modulus of the effective coupling frequency $\abs{\Omega_\text{eff}}/\Omega$ [upper (a) panel] and its phase $\arg(\Omega_\text{eff})$ (modulo $2\pi$) [lower (b) panel].  Dots are the results of the numerical calculations as discussed in the text, while the solid lines are  analytical predictions of the RW approximation. System parameters as in Fig.~\ref{fig:resonanceNORWA}.
	}
	\label{fig:Omegaeffettiva}
\end{figure}

Further insight on the effective coupling of the two pendulums, resulting from the temporal modulation, is given in Fig.~\ref{fig:Omegaeffettiva}. In panel (a) we present the numerically estimated magnitude of the effective coupling between the pendulums: this quantity is observable as the beat frequency in Fig.~\ref{fig:confrontoomegaeffnorwa}, or as the separation between peaks in the response spectra of Fig.~\ref{fig:resonanceNORWA}. Most remarkable features of Fig.~\ref{fig:Omegaeffettiva}(a) are the vanishing effective coupling at $I_0/w\simeq 1.2$ and $I_0/w\simeq 2.7$, corresponding to the zeros of only $\max(\abs{A_2})$ in Fig.~\ref{fig:Feffettiva}. In the following, we will see that these numerical values are related to the zeros of the zeroth-order Bessel function.

Another crucial consequence of the temporal modulation is shown in the lower panel Fig.~\ref{fig:Omegaeffettiva}(b). While the eigenmodes of the bare system correspond to in-phase and out-of-phase oscillations, the modulation allows to tune the relative phase of the oscillation of the two pendulums to any value from $0$ to $2\pi$. In mathematical terms, the effective coupling develops a non-trivial phase. Note how this phase displays $\pi$ jumps whenever the effective coupling goes through zero. Before proceeding, it is worth noting in Fig.~\ref{fig:resonanceNORWA} a sizeable global shift of the resonance curves towards lower frequencies for growing $I_0$. An explanation of this effect will be given in the next section.

\section{Analytical expressions within the rotating wave approximation}
Analytical insight in the physics of the modulated system can be obtained within the so-called rotating wave (RW) approximation, well-known from quantum optics. This approximation relies on assuming that the natural oscillation frequency $\omega_0$ is much larger than all other internal frequencies in the problem, that is $\omega_0\gg\max(\Omega, \gamma,w,\abs{\omega_{\rm ex}-\omega_0})$. 
Since the $\alpha$ variables rotate at $\approx \omega_0$ and their conjugate variables $\alpha^*$ rotate at $\approx -\omega_0$, the RW approximation is straightforwardly implemented by neglecting the $\alpha^*$ terms in the motion Equations~\eqref{nascostaalpha1noRWA} for the $\alpha$ variables.

To obtain an effective, temporally averaged form of the stroboscopic dynamics of the system, we introduce the new variables: $\beta_i(t)= \alpha_i(t)\, \exp[\ii \int_0^t v_i(t') \,\diff t'] \exp[\ii\, \omega_\text{ex} t]$. For the chosen sinusoidal form of the modulation, the phase factor involving the modulation is exactly equal to $1$ at the stroboscopic times $t_j=2\pi j/w$ that are considered in the figures. Since we are in the fast modulation regime $w\gg\Omega$, effective equations, which no longer depend on the time $t$, can be obtained~\cite{Dalibard_notes_CdF} by averaging the RW form of the equations \eqref{nascostaalpha1noRWA} over a modulation period $T=2\pi/w$:
\begin{align}
\dot{\beta}_1&=-\ii\, \left(\bar{\omega}_0-\omega_\text{ex}\right)\,\beta_1 -\gamma\,\beta_1+ \ii\, (\Omega_{12}^\text{eff}/2)\, \beta_2 +\ii f_\text{ex}^\text{eff}\label{beta1RWAex}\\
\dot{\beta}_2&=-\ii\, \left(\bar{\omega}_0-\omega_\text{ex}\right) \,\beta_2-\gamma\,\beta_2+\ii \,(\Omega_{21}^\text{eff}/2)\, \beta_1. \label{beta2RWAex}
\end{align}
The effective couplings in Eqs.~\eqref{beta1RWAex}-\eqref{beta2RWAex} are defined by:
\begin{equation*}
\Omega_{ij}^\text{eff}\,\equiv\, \frac{\Omega}{T}\int_0^T\!\!\diff t\,\e^{\ii \int_0^t  [v_i(t')-v_j(t')]\,\diff t'}
\end{equation*}
and they are complex conjugate to each other, $\Omega_{12}^\text{eff}=\Omega_{21}^{\text{eff}*}$. For the specific modulation considered here, the effective coupling has the simple expression:
\begin{equation}
\Omega_\text{eff}\equiv\, \Omega_{12}^\text{eff}=\Omega\,\e^{-2\ii I_0/w }\mathcal{J}_0\left(2 I_0/w\right)
\label{Omegaeff}
\end{equation}
in terms of the $\mathcal{J}_0$ zero-order Bessel function. \footnote{A non-zero phase in the driving \eqref{modulaz}, or equivalently a temporal shift in the stroboscopic sampling, will result in an extra phase factor in $\Omega_\text{eff}$ \cite{Sols08}.} The modulus and the phase of this quantity are plotted as a solid line in Fig.~\ref{fig:Omegaeffettiva}.  In particular, the modulus $\abs{\Omega_\text{eff}}$ shows a series of zeros, which are indicative of a complete {\em dynamical decoupling} between the two pendulums.

The same procedure must be applied to the amplitude of the external driving force in Eq.~\eqref{beta1RWAex}, which gives the effective averaged driving force:
\begin{equation*}
f_\text{ex}^{\text{eff}}\equiv\, \frac{f_\text{ex}}{T} \int_0^T\!\! \diff t\,\, \e^{\ii \int_0^t v_i(t') \, \diff t'}
\end{equation*}
that, for the specific modulation in Eq.~\eqref{modulaz}, has the form:
\begin{equation}
f_\text{ex}^{\text{eff}}=f_\text{ex} \, \e^{-\ii I_0/w} \mathcal{J}_0(I_0/w).
\label{fexteff}
\end{equation}
The zeros of $f_\text{ex}^{\text{eff}}$ determine the parameters at which complete {\em dynamical isolation} from the external force occurs.

Explicit forms for the steady oscillation regime can be derived by setting the time derivatives in Eqs.~\eqref{beta1RWAex}-\eqref{beta2RWAex} to zero. In this way, one obtains the following analytical form of the resonance curves as function of $\omega_\text{ex}$:
\begin{align}
\beta_1(\omega_\text{ex})= \frac{2\,f_\text{ex}^\text{eff}\,(\bar{\omega}_0 -\ii\gamma -\omega_\text{ex})
}{4(\bar{\omega}_0 -\ii \gamma-\omega_\text{ex})^2
-\abs{\Omega_\text{eff}}^2}\label{resonanceRWAeff1}\\
\beta_2(\omega_\text{ex})= \frac{f_\text{ex}^\text{eff} \,\Omega^*_\text{eff}}{4(\bar{\omega}_0 -\ii\gamma-\omega_\text{ex})^2-\abs{\Omega_\text{eff}}^2}. \label{resonanceRWAeff2}
\end{align}
While a qualitative agreement with the numerical predictions shown in Fig.~\ref{fig:resonanceNORWA} is already present at this level, there is still an overall global shift of the resonances. This shift is easily explained by including the leading order correction to the RW approximation. We allow for the $\alpha_i$ to also have a small counter-rotating part evolving at frequency $-\omega_\text{ex}$ by writing $\alpha_i(t)=\alpha_i^\text{rw}\,\e^{-\ii \omega_\text{ex} t}+ \delta \alpha_i^{\text{nrw}} \e^{\ii \omega_\text{ex} t}$. Since the $\alpha$ and the $\alpha^*$ are coupled by the non-RW terms in Eqs.~\eqref{nascostaalpha1noRWA} and their complex conjugate equations, co-rotating contributions appear in the equation for $\alpha$ from the counter-rotating terms of $\alpha^*$. A straightforward calculation gives, to the leading order: $\delta \alpha_i^{\text{nrw}}\simeq -{(\alpha_i^{\text{rw}})^*\,v_i(t)}/{(2\omega_0)}$.
After substituting this expression into Eq.~\eqref{nascostaalpha1noRWA} and isolating the non-rotating terms, by averaging over the period of the considered modulation, the effective frequency shift is:
\begin{equation}
\Delta\omega_0=-\langle v_i(t)^2\rangle_T/(2\omega_0) = - I_0^2/(4\omega_0).
\label{shift}
\end{equation}
This is the principal effect of the counter-rotating terms beyond the RW, and it is more and more important for growing $I_0$. All other non-RW contributions involving $\Omega$ and $\gamma$ are negligible for the chosen parameters. The shift can be interpreted as a classical analog of the quantum Bloch-Siegert shift of nuclear magnetic resonance \cite{BlochSiegert}. 

The shift Eq.~\eqref{shift} can be taken into account in Eqs.~\eqref{resonanceRWAeff1}-\eqref{resonanceRWAeff2} by replacing $\bar{\omega}_0$ with $\tilde{\omega}_0=\bar{\omega}_0+\Delta\omega_0$. In this way, an excellent agreement for the analytical spectra (solid lines) with the result of the numerical simulations (dots) is found in Fig.~\ref{fig:resonanceNORWA}. This suggests a way to extract an estimate of the effective coupling $\Omega_{\rm eff}$ from the numerical results. An explicit expression of it in terms of the oscillation amplitudes is obtained by taking the ratio of Eq.~\eqref{resonanceRWAeff1} and Eq.~\eqref{resonanceRWAeff2}:
\begin{equation}
\Omega_{\text{eff}}\,=2\, (\tilde{\omega}_0 +\ii\,\gamma-\omega_{\text{ex}}) \left(\beta_2/\beta_1\right)^*.
\label{clever_omegaeff}
\end{equation}
The result of replacing in Eq.~\eqref{clever_omegaeff} the $\beta_i$ with the numerically calculated $A_i$ is shown by the dots in Fig.~\ref{fig:Omegaeffettiva} and is compared with the analytical RW prediction of Eq.~\eqref{Omegaeff}. The agreement appears to be very good for both the magnitude and the phase of the coupling, in particular the position of the dynamical decoupling points at which $\Omega_{\text{eff}}=0$. The small discrepancies occur when both numerator and denominator of Eq.~\eqref{clever_omegaeff} go to zero and the procedure is more sensitive to numerical errors. 

A similar comparison for the effective driving force is performed in Fig.\ref{fig:Feffettiva}, where the
numerical results are compared to RW prediction of Eq.~\eqref{fexteff}. The agreement is again very good, in particular for what concerns the position of the dynamical isolation points for which $f_\text{ex}^\text{eff}=0$ and both $\max(\abs{A_{1,2}})=0$. Of course, the agreement gets worse when the inequality chain of Eq.~\eqref{ineq} is only marginally satisfied.

\section{Connection to the Bose-Hubbard model}
Before concluding, it is worth to highlight the direct connection of the RW description of the system of coupled pendulums to a driven-dissipative version of the Bose-Hubbard (BH) model of quantum condensed-matter. In the presence of an external gauge field~\cite{Dalibard_RMP_gauge}, the BH model is described by the Hamiltonian:
\begin{equation}
\ham=-\sum_{\ang{ij}} \left[J\,\e^{\ii\phi_{ij}}\hat{a}_i^\dagger \hat{a}_j+\textrm{h.c.}\right]+\sum_i \frac{U}{2}\hat{a}_i^\dagger\hat{a}_i^\dagger\hat{a}_i\hat{a}_i
\label{bh}
\end{equation}
where $\hat{a}_i$ and $\hat{a}_i^\dagger$ are bosonic on-site operators. The first term describes hopping of the bosonic particles: the hopping amplitude is $J$ and the phase $\phi_{ij}$ describes a non-trivial tunneling phase. Of course, non-rotating wave terms do not appear in the standard condensed-matter BH model Hamiltonian in Eq.~\eqref{bh} as they would correspond to processes where the total numbers of particles is not conserved.
If particles are instead injected from a coherent source and lost from the system, the theoretical description requires a driving term of the form $\ham_F= \sum_i[f_{\text{ex},i}(t)\,\hat{a}_i^\dagger + \textrm{h.c.}]$ as well as damping terms, to be typically included at the level of the Master Equation~\cite{Carusotto_Ciuti_RMP}. 

Under the classical approximation where operators are replaced by $\mathbb{C}$ numbers, equations of motion analogous to Eqs.~\eqref{resonanceRWAeff1}-\eqref{resonanceRWAeff2} are found. In particular, the complex hopping amplitude of the BH model corresponds to the complex coupling between the pendulums,
$J\,\e^{\ii\phi_{ij}} \leftrightarrow \Omega_{ij}^{\rm eff}$: the non-trivial Peierls phase $\phi_{ij}=\int_{r_{j}}^{r_{i}} \diff \mathbf{r} \cdot \mathbf{A}/\hbar$ describing the effect of an external vector potential acting on the quantum particles~\cite{Dalibard_RMP_gauge} corresponds to a non-trivial phase of the $\Omega_{ij}^{\rm eff}$ coupling. An on-site interaction term analogous to the $U$ term in Eq.~\eqref{bh} directly appears when the anharmonicity of the pendulums is taken into account beyond the linearized motion Equations \eqref{eq1}-\eqref{eq4}.

The modulation scheme that has been envisaged in the present Letter for coupled pendulums corresponds to a temporal modulation of the on-site energies of the BH lattice. There is however one crucial difference worth noting: while the global shaking of the optical lattice potential that is typically used for this purpose in ultracold atom experiments~\cite{Madison,Lignier,Zenesini09,Struck2012} can only provide modulation amplitudes that are linearly dependent on the site position~\cite{Dalibard_notes_CdF}, systems of pendulums allow for an individual addressing of each single pendulum. This freedom will be very useful in view of generating synthetic gauge field configurations in our mechanical system.

\section{Final remarks and conclusions}
Inspired by analogous quantum effects, in this Letter we have theoretically studied dynamical localization and dynamical isolation effects in a classical system of two coupled pendulums when a temporally periodic modulation of the oscillation frequencies is applied to them. The dynamical decoupling effect can be used in mechanical engineering to suppress the coupling between the two pendulums, while the dynamical isolation allows to isolate the system from external forces.
Of course, our results are valid for coupled oscillators of any physical nature with a time dependent frequency, for instance temporally modulated RLC electric circuits.

From the point of view of fundamental physics, the non-trivial coupling phase between the pendulums is analogous to the Peierls phase of Bose-Hubbard models of quantum condensed-matter physics. In analogy to orbital magnetism and topological insulation, new intriguing phenomena are expected to appear in multi-dimensional lattices of many temporally modulated pendulums. Further exciting developments in nonlinear physics are expected to arise as a result of the intrinsic anharmonicity of pendulums.

\acknowledgments
We are grateful to Nicola Pugno, Andr\'e Eckardt and Jean Dalibard for stimulating discussions. We acknowledge partial financial support from ERC via the QGBE grant and from Provincia Autonoma di Trento.


\begin{thebibliography}{0}
\bibitem{CTH}
	\Name{Haroche S., Cohen-Tannoudji C., Audoin C. \and Schermann J. P.}
	\REVIEW{Phys. Rev. Lett.}{24}{1970}{861}.

\bibitem{Dunlap}
	\Name{Dunlap D. H. \and Kenkre V. M.}
	\REVIEW{Phys. Rev. B}{34}{1986}{3625}.

\bibitem{Ignatov}
	\Name{Ignatov A. A. \and Romanov Y. A.}
	\REVIEW{Phys. Status Solidi B}{73}{1976}{327}.

\bibitem{Zhao}
	\Name{Zhao X.-G.}
	\REVIEW{J. Phys.}{6}{1994}{2751}.

\bibitem{Holthaus}
	\Name{Holthaus M.}
	\REVIEW{Phys. Rev. Lett.}{69}{1992}{351}.
		
\bibitem{Holthaus2}
	\Name{Holthaus M. \and Hone D.}
	\REVIEW{Phys. Rev. B}{47}{1993}{6499}.
	
\bibitem{Grossmann}
	\Name{Gro{\ss}mann F., Dittrich T., P. Jung \and H\"{a}nggi}
	\REVIEW{Phys. Rev. Lett.}{67}{1991}{516}.

\bibitem{Grifoni}
	\Name{Grifoni M. \and H\"{a}nggi P.}
	\REVIEW{Phys. Rep.}{304}{1998}{229}.   	
	
\bibitem{Madison}
	\Name{Madison K. W., Fischer M. C., Diener R. B., Niu Q. \and Raizen M. G.}
	\REVIEW{Phys. Rev. Lett.}{81}{1998}{5093}.

\bibitem{DLvsCDT}
	\Name{Kayanuma Y. \and Saito K.}
	\REVIEW{Phys. Rev. A}{77}{2008}{010101}.

\bibitem{Lignier}
	\Name{Lignier H., Sias C., Ciampini D., Singh Y., Zenesini A., Morsch O. \and Arimondo E.}
	\REVIEW{Phys. Rev. Lett.}{99}{2007}{220403}.

\bibitem{Arimondo09}
	\Name{Eckardt A., Holthaus M., Lignier H., Zenesini A., Ciampini D., Morsch O. \and Arimondo E.}
	\REVIEW{Phys. Rev. A}{79}{2009}{013611}.

\bibitem{EckardtHolthaus}
	\Name{ Eckardt A., Weiss C. \and Holthaus M.}
	\REVIEW{Phys. Rev. Lett.}{95}{2005}{260404}.
   	
\bibitem{EckardtHolthausEPL}
	\Name{Eckardt A. \and Holthaus M.}
	\REVIEW{Europhys. Lett.}{80}{2007}{50004}.

\bibitem{Zenesini09}
	\Name{Zenesini A., Lignier H., Ciampini D., Morsch O. \and Arimondo E.}
	\REVIEW{Phys. Rev. Lett.}{102}{2009}{100403}.

\bibitem{Struck2012}
	\Name{Struck J., \"{O}lschl\"{a}ger C., Weinberg M., Hauke P., Simonet J., Eckardt A., Lewenstein M., Sengstock K. \and Windpassinger P.}
	\REVIEW{Phys. Rev. Lett.}{108}{2012}{225304}.

\bibitem{Sols08}
	\Name{Creffield C. E. \and Sols F.}
	\REVIEW{Phys. Rev. Lett.}{100}{2008}{250402}.

\bibitem{Kolovsky}
	\Name{Kolovsky A. R.}
	\REVIEW{Europhys. Lett.}{93}{2011}{20003}.

\bibitem{CommentKolovsky}
	\Name{Creffield C. E. \and Sols F.}
	\REVIEW{Europhys. Lett.}{101}{2013}{40001}.

\bibitem{Hauke}
	\Name{Hauke P., Tieleman O., Celi A., \"{O}lschl\"{a}nger C., Simonet J., Struck J., Weinberg M., Windpassinger P., Sengstock K., Lewenstein M. \and Eckardt A.}
	\REVIEW{Phys. Rev. Lett.}{109}{2012}{145301}.

\bibitem{Struck2013}
	\Name{Struck J., Weinberg M., \"{O}lschl\"{a}nger C., Windpassinger P, Simonet J., Sengstock K., H\"{o}ppner R., Hauke P., Eckardt A., Lewenstein M. \and Mathey L.}
	\REVIEW{Nat. Phys.}{9}{2013}{738}.
	
\bibitem{Aidelsburger}
	\Name{Aidelsburger M., Atala M., Lohse M., Barreiro J. T., Paredes B. \and Bloch I.} 
   	\REVIEW{Phys. Rev. Lett.}{111}{2013}{185301}.

\bibitem{Miyake}
	\Name{Miyake H., Siviloglou G. A., Kennedy C. J., Burton W. C. \and Ketterle W.}
   	\REVIEW{Phys. Rev. Lett.}{111}{2013}{185302}.
   	 	
\bibitem{Longhi}
	\Name{Longhi S., Marangoni M., Lobino M., Ramponi R., Laporta P., Cianci E. \and Foglietti V.}
  \REVIEW{Phys. Rev. Lett.}{96}{2006}{243901}.	
   	
\bibitem{Rechtsman}
	\Name{Rechtsman M. C., Zeuner J. M., Plotnik Y., Lumer Y., Podolsky D., Dreisow F., Nolte S., Segev M. \and Szameit A.}
	\REVIEW{Nat.}{496}{2013}{196}.
	
\bibitem{Hafezi}
	\Name{Hafezi M., Mittal S., Fan J., Migdall A. \and Taylor J. M.}
	\REVIEW{Nat.Phot.}{7}{2013}{1001}.

\bibitem{Butikov}
	\Name{Butikov E. I.}
	\REVIEW{Am. J. Phys.}{69}{2001}{755}.

\bibitem{Arnold}
	\Name{Arnold V.I.}   
   	\Book{Mathematical Methods of Classical Mechanics}
   	\Publ{Springer-Verlag, New York}
   	\Year{1989}. 

\bibitem{Berthet}
	\Name{Berthet R., Petrosyan A. \and Roman B.}
	\REVIEW{Am. J. Phys.}{70}{2002}{774}.
	
\bibitem{Dalibard_notes_CdF}
	\Name{Dalibard J.}
	{\em Lecture Notes at Coll\`ege de France, A.Y. 2012-13 \url{www.phys.ens.fr/~dalibard/CdF/Cours_2013.pdf}}

\bibitem{BlochSiegert}
	\Name{Bloch F. \and Siegert A.}
	\REVIEW{Phys. Rev.}{57}{1940}{522}.

\bibitem{Dalibard_RMP_gauge}
	\Name{Dalibard J., Gerbier F., Juzeli\={u}nas G. \and \"Ohberg P.}
	\REVIEW{Rev. Mod. Phys.}{83}{2011}{1523}.	

\bibitem{Carusotto_Ciuti_RMP}
	\Name{Carusotto I. \and Ciuti C.}
	\REVIEW{Rev. Mod. Phys.}{85}{2013}{299}.

\end{thebibliography}
\end{document}